\documentclass[amsmath,amssymb,aps,
showpacs]{revtex4}

\usepackage{graphicx}
\usepackage{dcolumn}
\usepackage{subfigure,soul}
\usepackage{bm}

\begin{document}

\title[]{Anisotropic Dirac cones in monoatomic hexagonal lattices. A DFT study}

\author{A. M. Rojas-Cuervo, K. M. Fonseca-Romero and R. R. Rey-Gonz\'alez\footnote{Corresponding author: rrreyg@unal.edu.co.}}

\address{ Departamento de F\'isica, Universidad Nacional de Colombia, Ciudad Universitaria, C.P. 111321, Bogot\'a D.C. Colombia.}%

\begin{abstract}
In the last few years, the fascinating properties of graphene have been thoroughly investigated.
The existence of Dirac cones is the most important characteristic of the electronic band-structure of graphene.
In this theoretical paper, hexagonal monolayers of silicon (h-Si) and germanium (h-Ge) are examined using density functional theory, within the generalized gradient approximation.
Our numerical results indicate that both h-Si and h-Ge are chemically stable.
The lattice parameters, electronic dispersion relations and densities of states for these systems are reported.
The electronic dispersion relations display Dirac cones with the symmetry of an equilateral triangle (the group  D$_3$) in the vicinity of the K points.
Hence, the Fermi velocity depends on the wave vector direction around $K$ points.
Fermi velocities for holes and electrons are significantly different.
The maximum and minimum Fermi velocities are also reported.
\end{abstract}

\pacs{
{31.15.E-}, 
{61.48.Gh}, 
{71.15.Dx}, 
{73.22.-f} 
}
\maketitle

\section{\label{sec:level1}INTRODUCTION}

One of the goals of nanotechnology is the development of new materials by controlled manipulation at the nanometer scale \cite{IntroNano}.
Perhaps the most important new material, recently discovered, is graphene  \cite{Novoselov22102004}.
This monolayer of carbon atoms arranged in a regular hexagonal pattern, was first isolated by mechanical exfoliation of graphite \cite{Novoselov22102004}.
The isolation of graphene was considered for many years to be an impossible task, because perfect free-standing 2D crystals can not exist \cite{Peierls,Landau,Mermin}, due to their thermodynamic instability.
The stability of experimentally produced samples of graphene has been ascribed to height fluctuations \cite{Fasolino,Meyer}.

The unique properties of graphene (strength, thermal and electrical conductivity, transparency) have been studied theoretically and experimentally \cite{Novoselov22102004, Novoselov2005, Novoselov2007}.
Graphene is a zero-gap material, as was early noticed by Wallace \cite{Wallace}.
Near the $K$-point (in the reciprocal lattice), where the conduction and valence bands touch each other, the energy grows linearly with the momentum.
In this zone, known as Dirac cone, electrons and holes behave as massless particles with an effective speed $c^*\approx 10^6$ m/s \cite{Wallace}.
In this work we consider Si (Silicon) and Ge (Germanium), which are usually employed in technological applications, and we investigate if they can be arranged in a stable 2D honeycomb pattern.
Our numerical simulations show not only that h-Si, h-Ge are indeed stable, but also that they exhibit Dirac cones.

Some details of the numerical method employed in this paper are introduced in section \ref{sec:level2}.
Our numerical findings for h-Si, h-Ge and h-C (used for comparison), are described in section \ref{sec:level3a}.
There, the lattice parameters, the dispersion relations, the densities of states (DOS) and the Fermi velocities of electrons and holes, are reported.
Our conclusions are summarized in the last section (\ref{sec:level5}).

\section{\label{sec:level2}COMPUTATIONAL DETAILS}

\textit{Ab initio} methods, especially the Density Functional Theory (DFT), have been very useful for the theoretical investigation of condensed-matter systems.
In particular, the SIESTA (\textit{Spanish Initiative for Electronic Simulations with Thousands of Atoms}) code implementation of DFT \cite{J.Phys.Condens.Matter.142002} is employed in academic \cite{reich}, industrial \cite{Motorola} and applied research \cite{investigacion} of solid-states systems ranging from molecules to infinite periodic 3D systems.
The SIESTA code allows efficient computation of electronic structure, vibrational properties and geometry optimization.
Since SIESTA implements algorithms which scale linearly with the number of atoms, it enables accurate calculations with relatively small computer facilities.

We have used the SIESTA code within the generalized gradient approximation (GGA), considering Perdew-Burke-Ernzerhof (PBE) \cite{PhysRevLett.77.3865} exchange and correlation functionals, to study the electronic and structural properties of h-C, h-Si and h-Ge monolayers.
Valence electrons are described using localized numerical atomic orbitals with a double$-\zeta$ polarized (DZP) basis set (taking into account spin effects).
Core electrons are implicitly treated with non-local norm-conserving pseudopotentials \cite{PhysRevB.43.1993, ATOM} and optimized cut off radii $r_c$ \cite{CamiloEspejo,ClaudiaBarrera}.
These radii were determined by imposing a maximum energy difference of $1~mRy$ on the transferability test \cite{siesta-manual, ATOM}.
An energy cutoff of $125~Ry$ ($1700~eV$) was used for real space integrations.
A relaxation process, in which the atoms are moved step by step, are used to assess the structural stability of h-C, h-Si and h-Ge monolayers.
In each step the force on the atoms is minimized and the pseudopotential is calculated iteratively.
We consider that the chemical structure is stable if the maximum atomic force is smaller than  $40~meV\cdot$\r{A} \cite{siesta-manual}.
Additionally, the cohesive energies are calculated in each case.

For simulation purposes, the infinite honeycomb single layer studied is set to coincide with the $xy$ plane.
Due to the finite size of the 3D primitive cell used, our results would correspond to an infinite set of interacting parallel layers, unless ${a}_z,$ the component of the primitive vector along the $z$-direction, is chosen to be large enough to preclude interaction between the layers.
Here, we used  ${a}_z=100a$, where $a$ is the 2D lattice constant.
The buckling of h-Si and h-Ge\cite{Takeda1994,Cahangirov2009}, not taken into account in this study, could enhance the anisotropy of Dirac cones reported in this manuscript.

\section{\label{sec:level3a}RESULTS}
\subsection{\label{sec:level3}LATTICE PARAMETERS}
The relaxation method described in the previous section converges for h-Si and h-Ge (and also for graphene).
In order to assess the stability of the systems studied in this paper, their cohesive energies are calculated (see Table \ref{table0}).
Our calculations for bulk cohesive energies are in good agreement with well known values \cite{kittel} and are slightly larger than the corresponding values for the 2D structures.


\begin{table}[ht!]
\caption{\label{table0} Values of cohesive energy including spin effects.}
\centering
\begin{tabular}{l c c c c c c}\hline\hline
Materials& h-C & Bulk C & h-Si & Bulk Si & h-Ge & Bulk Ge\\ \hline
Cohesive energy \scriptsize{(eV/atom)}& 7.63 & 7.64 & 3.90 & 4.62 & 3.08 & 3.81\\ \hline\hline
\end{tabular}
\end{table}
We therefore conclude that all of them are chemically stable.
Data found for bond length ($a_{0}$), lattice constant ($a$), and actual maximum atomic force tolerance ($f_{tol}$) for each considered monolayer are reported in Table \ref{tab:table1}.

\begin{table}[ht!]
\caption{\label{tab:table1} Calculated $a_{0}$, $a$, and $f_{tol}$ of hexagonal monolayers including spin effects.}
\small
\centering
\begin{tabular}{l l l l}\hline\hline
Monolayer &$a_{0}$ \scriptsize{(\r{A})}&$a$ \scriptsize{(\r{A})}& $f_{tol}$ \scriptsize{(eV\r{A}$^{-1}$)}\\ \hline
h-C &1.419&2.458&0.002 \\
h-Si&2.230&3.860&0.2$\times 10^{-3}$\\
h-Ge&2.310&4.001&0.030 \\\hline\hline
\end{tabular}
\end{table}

Although the results for graphene are well known, we have included them in Table \ref{tab:table1} for comparison.
They are in very good agreement with previous theoretical and experimental reports \cite{Diaye,Dedkov,Goncalves}.  
Recent experiments have shown that Silicon monolayers are stable.
However, the reported lattice parameter $a$ seems to depend on the substrate used in its growth: {\normalsize$a=$ 3.3  \r{A}} \cite{lalmi:223109}, {\normalsize$a=$ 3.65 \r{A}} \cite{PhysRevLett.108.245501} and {\normalsize$a=$3.88 \r{A}} \cite{padova:261905} for Silicon monolayers grown on {\normalsize$A_{g}(111)$}, {\normalsize$A_{g}(110)$} and {\normalsize$(0001)-ZrB_{2}$}, respectively.
When these results are compared with ours, differences are {\normalsize 14.55$\%$, 5.49$\%$} and {\normalsize0.46$\%$}, respectively.
To the best of our knowledge, there are no experimental results for the lattice parameter of h-Ge and previous theoretical
reports run from  {\normalsize$a=$ 3.97 \r{A}} \cite{PhysRevB.80.155453}, {\normalsize$a=$ 4.001 \r{A}} \cite{Suzuki20102820},
{\normalsize$a=$ 4.126 \r{A}} \cite{PhysRevB.79.115409},
as far as {\normalsize$a=$ 4.127 \r{A}} \cite{doi:10.1021/jp203657w}.

\subsection{\label{sec:level4}DISPERSION RELATIONS AND DENSITY OF STATES}
\begin{figure}[ht!]
\centering
\includegraphics[scale=0.325]{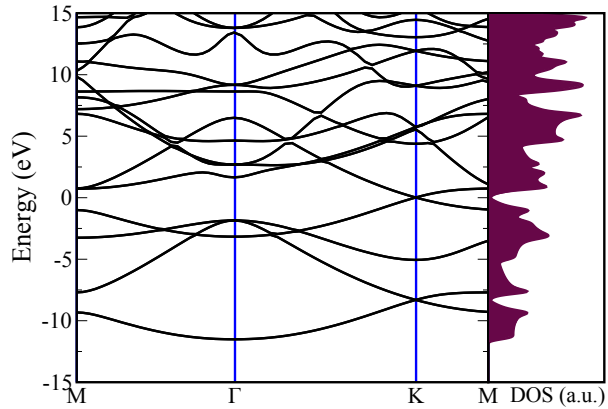}
\caption{\small Electronic dispersion relation (left panel) and DOS (right panel) for h-Si.}
 \label{figBSipbs}
\end{figure}

\begin{figure}[ht!]
\centering
\includegraphics[scale=0.325]{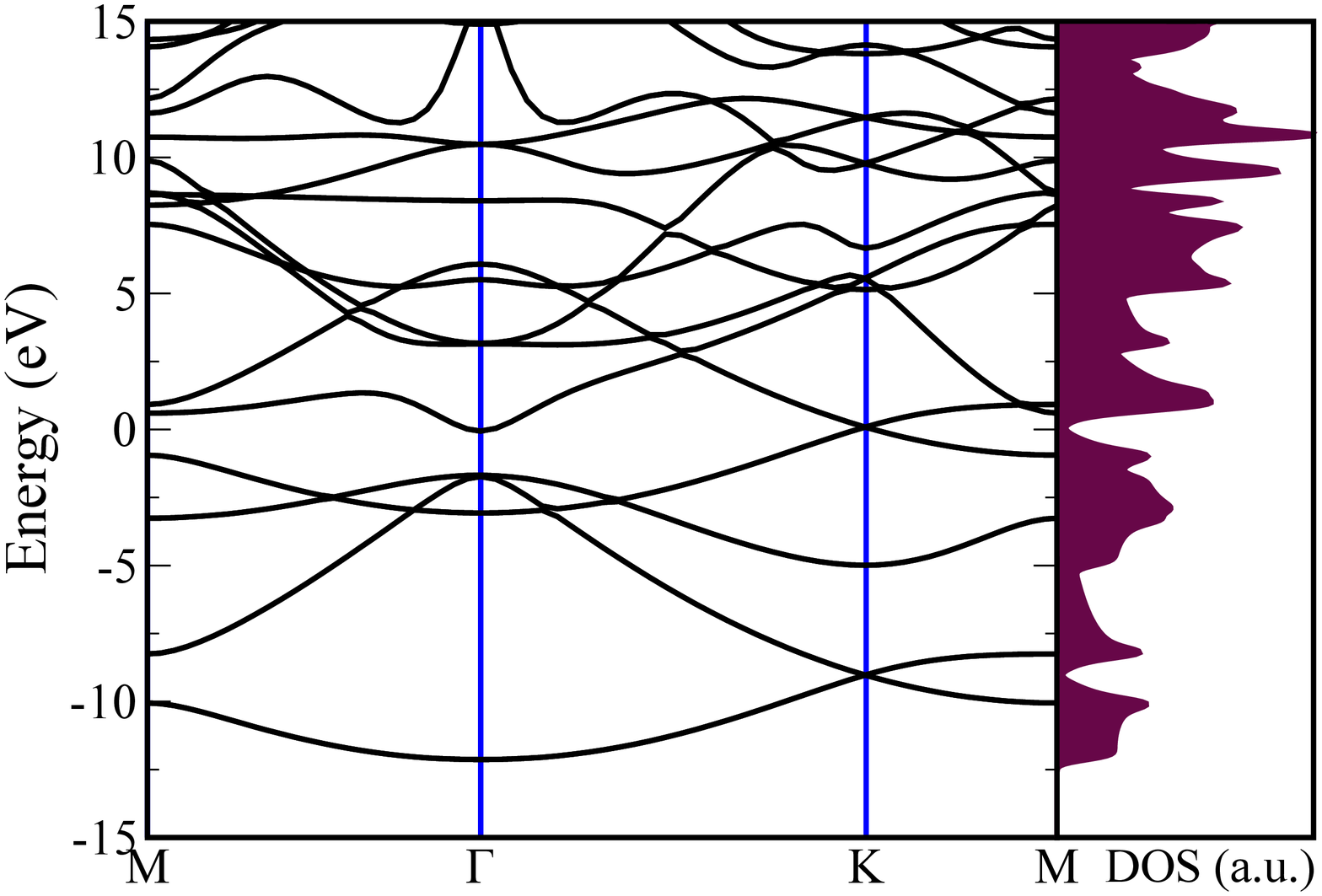}
\caption{\small Electronic dispersion relation (left panel) and DOS (right panel) for h-Ge.}
 \label{figBGepbs}
\end{figure}

The electronic dispersion relations corresponding to the stable structures of h-Si, h-Ge and graphene were calculated.
They turned out to be very different to those of their 3D diamond crystals.
While the 3D crystals are semiconductors of indirect gap of $1.12~eV$ (Si) and $0.67~eV$ (Ge) at $T=300~K$ \cite{davies,Ashcroft}, our results show that the corresponding 2D monolayers are semimetals.
Indeed, as the dispersion relations and DOS show (see Figs. \ref{figBSipbs} and \ref{figBGepbs}), there is no energy gap between valence and conduction bands.
Both bands intersect at two inequivalent points K and K$^\prime$ in the reciprocal space.
These conclusions hold even if spin effects are ignored.
We make a comparison between our numerical findings and results previously reported in the literature in Table \ref{Cgaps}.
Our numerical results agree with some previous results but are at variance with others.

\begin{table}[ht!]
\caption{Energy gap values for h-Si and h-Ge, in eV.} \label{Cgaps}
{\small
\centering
 \begin{tabular}{lcccccccc}\hline\hline
System & This work$^{a}$ & Ref. \cite{PhysRevB.80.155453}$^{b}$ & Ref. \cite{doi:10.1021/jp203657w}$^{a}$  & Ref. \cite{houssa:082111}$^{b}$ &
Ref. \cite{PhysRevB.76.075131}$^{d}$ & Ref. \cite{PhysRevB.79.115409}$^{a,b}$   & Ref. \cite{Suzuki20102820}$^{c}$ &
Ref. \cite{rojas-cuervo:169}$^{a}$  \\\hline
h-Si&Semimetal  & Semimetal& Metal    & -     & Metal,$Eg\neq0$  &  Semimetal      &  0.064(Direct)    & Semimetal\\
h-Ge& Semimetal &  Metal   & Metal & Metal & -                & Semimetal       & -0.444(Semimetal) & Semimetal\\\hline\hline
\end{tabular}\\
$^{a}$ GGA,  $^{b}$ LDA, $^{c}$ LSDA, $^{d}$ Tight Binding.
}
\end{table}

\begin{figure}[ht!]
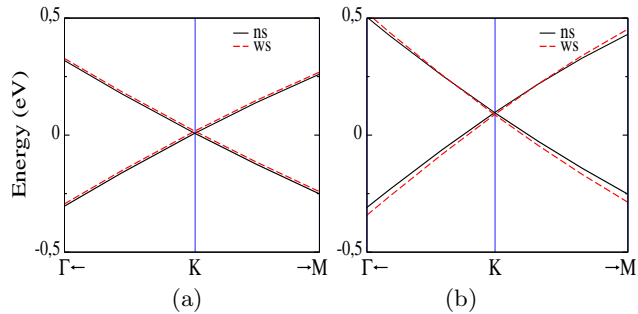

\centering
\includegraphics[width=4.2cm, height=3.6cm]{figure6a-1}
\includegraphics[width=4.0cm, height=3.6cm]{figure7a-1}
\centerline {(a) \hspace{3.0cm} (b)}
\caption{\small Electronic dispersion relations for (a) h-Si and (b) h-Ge in the vicinity of their $K$-points.
Black solid lines (ns) correspond to results ignoring spin effects; red dashed lines (ws) corresponds to results which take spin effects into account.}
 \label{figPDiracSiGe}
\end{figure}

\begin{figure}[ht!]
\centering
\includegraphics[scale=0.2]{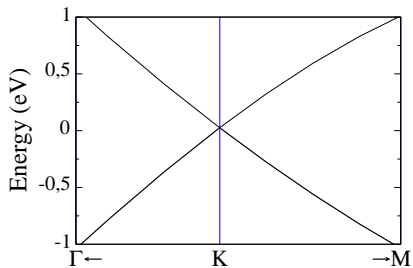}
\caption{\small Electronic dispersion relation for graphene near the $K$-point.}
 \label{figPDiracC}
\end{figure}

Figs.  \ref{figPDiracSiGe}(a) and \ref{figPDiracSiGe}(b) display magnifications of the electronic dispersion relations of
 h-Si and h-Ge around their respective K-points.
These dispersion relations are similar to that of graphene in the neighborhood of its K-point (see Fig. \ref{figPDiracC}).
Indeed, around their K-points, the dispersion relations of  h-Si, h-Ge and graphene are linear and resemble the dispersion
relation of relativistic electrons.
Thereby, simulations of (2+1)-dimensional electrodynamics involving massless fermions could be carried out in all of these
layer materials.
Thus, the existence of Dirac cones is not exclusive of graphene, but it is shared by other monoatomic two-dimensional hexagonal
lattices such as h-Si and h-Ge.
However, not all hexagonal layers exhibit Dirac cones.
Our theoretical results indicate that diatomic hexagonal monolayers do not present this behavior \cite{rojas-rey-2}.

Thus far, our numerical results are in good agreement with the predictions of a simple tight-binding model \cite{RevModPhys.81.109}.
However, our results also hint to physics beyond the tight-binding model, which anticipates circular surfaces of constant energy around the Dirac point \cite{Wallace} and electron-hole symmetry \cite{RevModPhys.81.109}.
In fact, our results for the electronic dispersion relations of h-Si, h-Ge and graphene near their Dirac points show two types of asymmetry, which have been overlooked in previous theoretical studies.
The first type of asymmetry distinguishes between electron and holes.
Experimental electron-hole asymmetries in graphene have been reported by Kai-Chieh Chuang {\it et al} \cite{Chuang} and by Jiam Xue {\it et al} \cite{Xue}.
The second type of asymmetry, which is larger than the first, distinguishes the direction of the wavevector.

Both types of asymmetry are present whether spin effects are taken into account or not.
Spin brings about energy corrections, as can be seen in Figs.  \ref{figPDiracSiGe}(a) and \ref{figPDiracSiGe}(b).
In order to quantify the above-mentioned asymmetries, as well as the relative importance of spin effects, we calculate the corresponding Fermi velocities (see Table \ref{tab:tableVC}).
For this purpose, a linear regression is performed, starting at the $K$-point and ending at points $\Gamma$  and  $M.$
Spin induces variations of Fermi velocities ranging from 0.1\%, for electron velocity of graphene in the direction of K-M, and up to $15\%$, for electron velocity of h-Si in the direction of K-M, as shown on the third line of each subpanel in the table.
These variations in electron (or hole) velocity can be positive or negative, and can magnify or reduce the asymmetries introduced in the previous paragraph.

Electron-hole symmetry is broken.
Electron and hole Fermi speeds along the same direction can differ for as little as 0.2\% (K-M direction in graphene) or for as much as 8.4\% (K-M direction in h-Ge).
Electron-hole asymmetry in graphene is magnified by spin effects because they reduce electron speed and increase hole speed.
This asymmetry in h-Ge is also magnified by spin effects.
Although spin always increases Fermi speeds, its effect on holes is larger than on electrons.

Finally, Fermi speed for electrons and for holes depends on its direction.
Electron and hole Fermi speeds along the directions K-$\Gamma$ and K-M are given in Table \ref{tab:tableVC}.
As will be clear below in the text, the smallest speed occurs along the K-M direction and the largest along the K-$\Gamma$ direction.
The relative variation of electron Fermi speed  between the two analyzed directions is of $36.6\%$ for graphene, of $45.9\%$ for h-Ge and of $61,4\%$ for h-Si.
Similar differences are found in hole velocities: $28.7\%$ in graphene,  $57.5\%$ in h-Si, and  $48.5\%$ in h-Ge.

Experimentally measured electron speeds show large discrepancies.
The following electron Fermi velocities of graphene have been  reported: $0.79\times10^{6}~m/s$ \cite{Guohong-Li}, $1.093\times10^{6}~m/s$ \cite{Chuang}, $1.10\times10^{6}~m/s$ \cite{Siegel}, and $1.16\pm0.01\times10^{6}~m/s$ \cite{Xue}.
The only published measure for hole Fermi speed of graphene, that we are aware of,  gives the value $0.94\pm0.02\times10^{6}~m/s$ \cite{Xue}.
This value, and one of the experimental reports of electron Fermi speeds, are below the range predicted by our numerical approach.
The other three experimental values fall within the range set by our approach.
Fermi speed of electrons and holes in h-Si  of $1.3\times10^{6}~m/s$, measured by Vogt \emph{et al.} \cite{Vogt}, is consistent with our results.
For h-Ge, as far as we know, there are no experimental reports of Fermi speed.
We speculate that the anisotropy of Fermi speed is one of the reasons behind the discrepancy of the measured electron speed of graphene.

\begin{table*}[ht!]
\caption{\label{tab:tableVC}Electron and hole Fermi speeds $v_{f}$ (x$10^{6}m/s$) of graphene (h-C), h-Si and h-Ge as a function of their direction, and ignoring (ns) or taking into account spin effects (ws). $\Delta$ is the percentual variation of the speed due to spin.}	
\small
\centering
\begin{tabular}{c c |c |c ||c |c }  \hline \hline
& \multicolumn{3}{c}{$\textbf{electrons}$} &  \multicolumn{2}{c}{$\textbf{holes}$}\\ \hline
 \rule{0pt}{3.5mm}
& & $v_{f}$ ($K\rightarrow\Gamma$)&$v_{f}$ ($K\rightarrow M$)&$v_{f}$ ($K\rightarrow\Gamma$) &$v_{f}$ ($K \rightarrow M$) \\ \hline
&ns&1.406& 0.998& 1.409& 0.996\\
\textbf{h-C}&ws&1.365& 0.999&1.385&1.076\\
&$\Delta$(\%) & -3.0\%& 0.1\%&-1.73\%&7.4\% \\ \hline
&ns&1.478&1.078&1.550&1.062\\
\textbf{h-Si}&ws&1.509&0.935&1.533&0.973\\
&$\Delta$(\%) & 2.0\%& -15.3\%&-1.1\%&-9.1\% \\ \hline
&ns&1.644&1.047&1.569&0.981\\
\textbf{h-Ge} &ws&1.746&1.197&1.639&1.104\\
&$\Delta$(\%) & 5.8\%& 12.5\%&4.3\% &11.1\% \\ \hline \hline
\end{tabular}
\end{table*}

\begin{figure}[ht!]
\centering
\subfigure[]{\includegraphics[width=4.0cm,height=4.6cm]{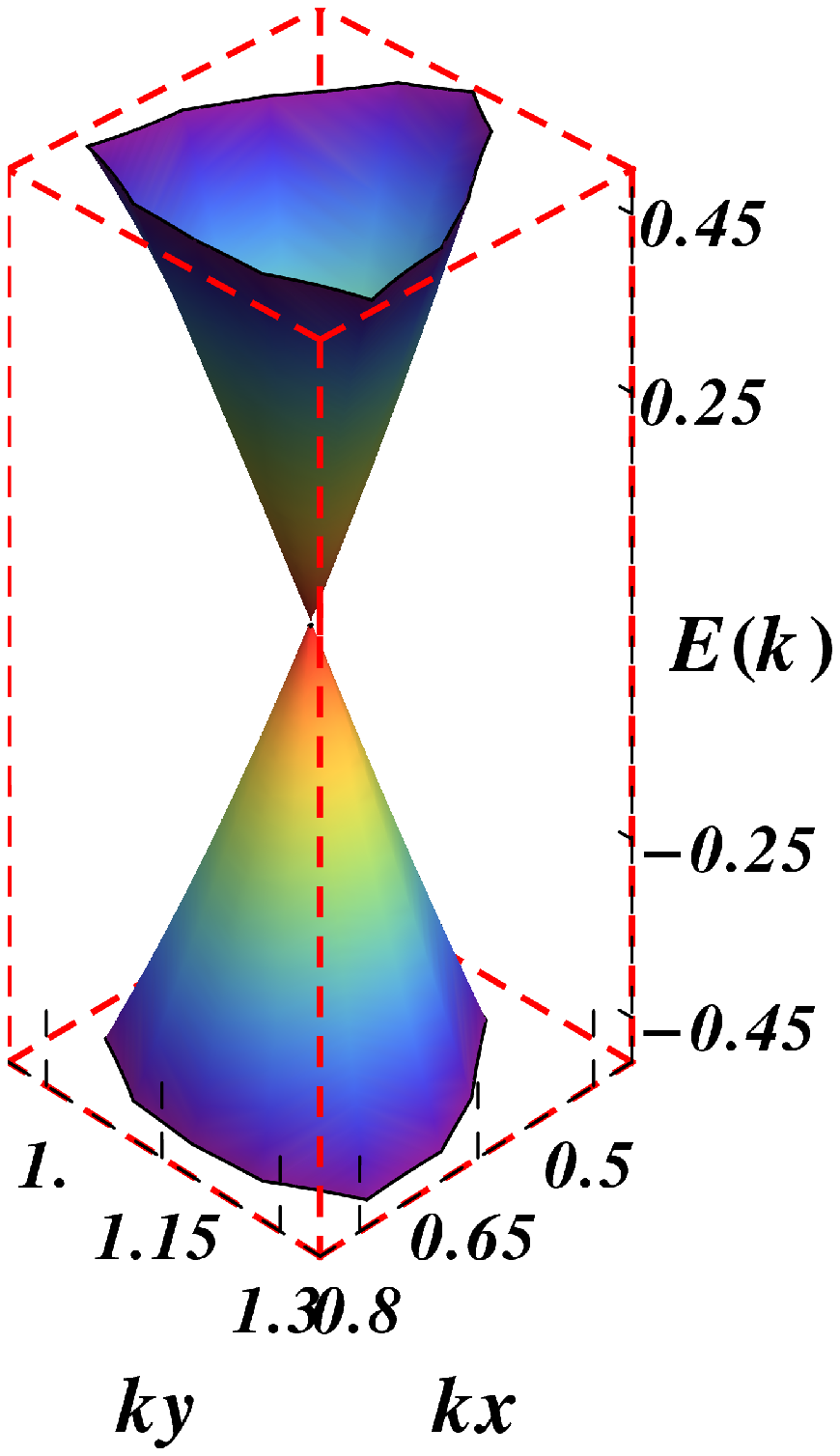}}\hspace{0.5cm}
\subfigure[]{\includegraphics[width=4.0cm,height=4.6cm]{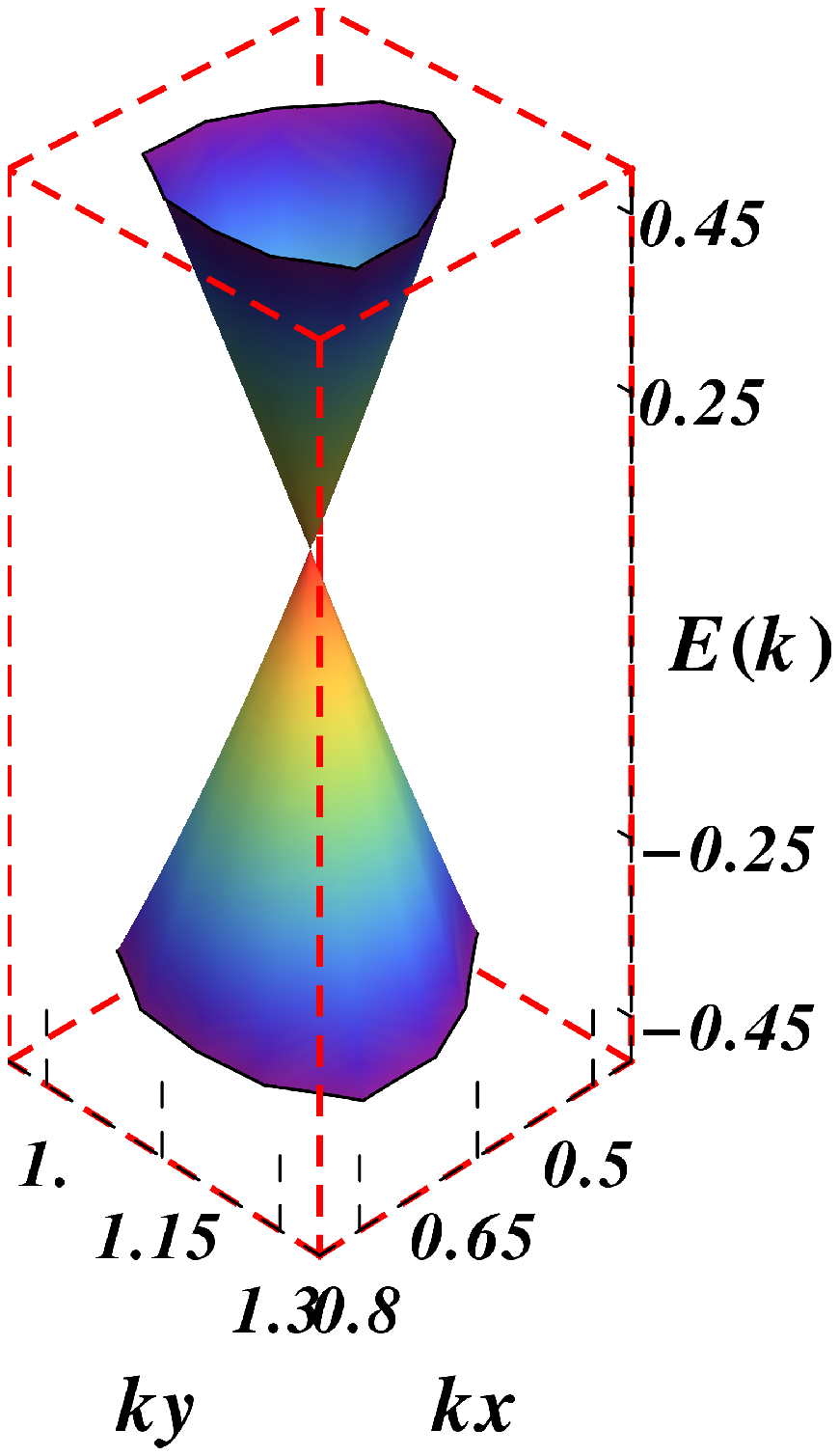}}\\
\caption{\small Electron and hole Dirac cones, around the $K$-point at $(\frac{2}{3},\frac{2\sqrt{3}}{3})$ for (a) h-Si and (b) h-Ge. Color online}
\label{DiracCones3D}
\end{figure}

In order to better explore the direction dependency of Fermi speeds and the electron-hole asymmetry, we plot the Dirac cones of h-Si (Fig \ref{DiracCones3D}(a)) and of h-Ge (Fig \ref{DiracCones3D}(b)).
Dirac cones associated with holes are shorter that Dirac cones for electrons, especially for h-Ge.
Therefore, the approximation of massless Dirac Fermions for holes, when compared with electrons, is valid in a smaller region around the K-point.
This observation can be important in experiments involving magnetic fields or phonon-assisted transitions.

\begin{figure}[h!]
\centering
\subfigure[]{\includegraphics[width=3.2cm,height=6.5cm]{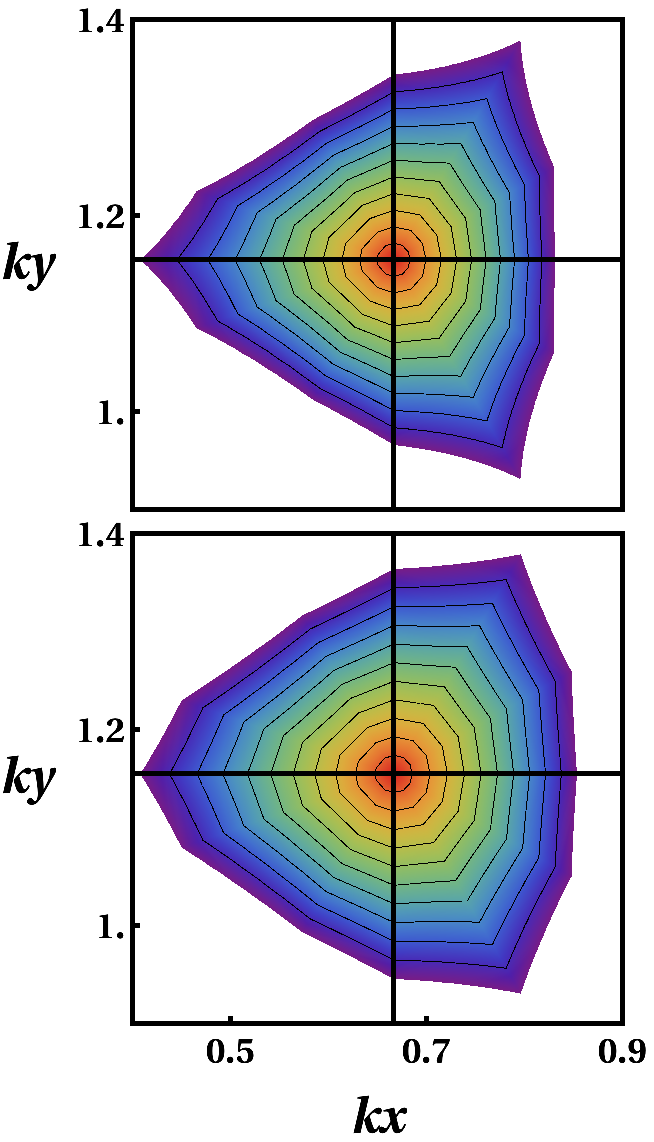}}\hspace{0.3cm}
\subfigure[]{\includegraphics[width=4.6cm,height=6.5cm]{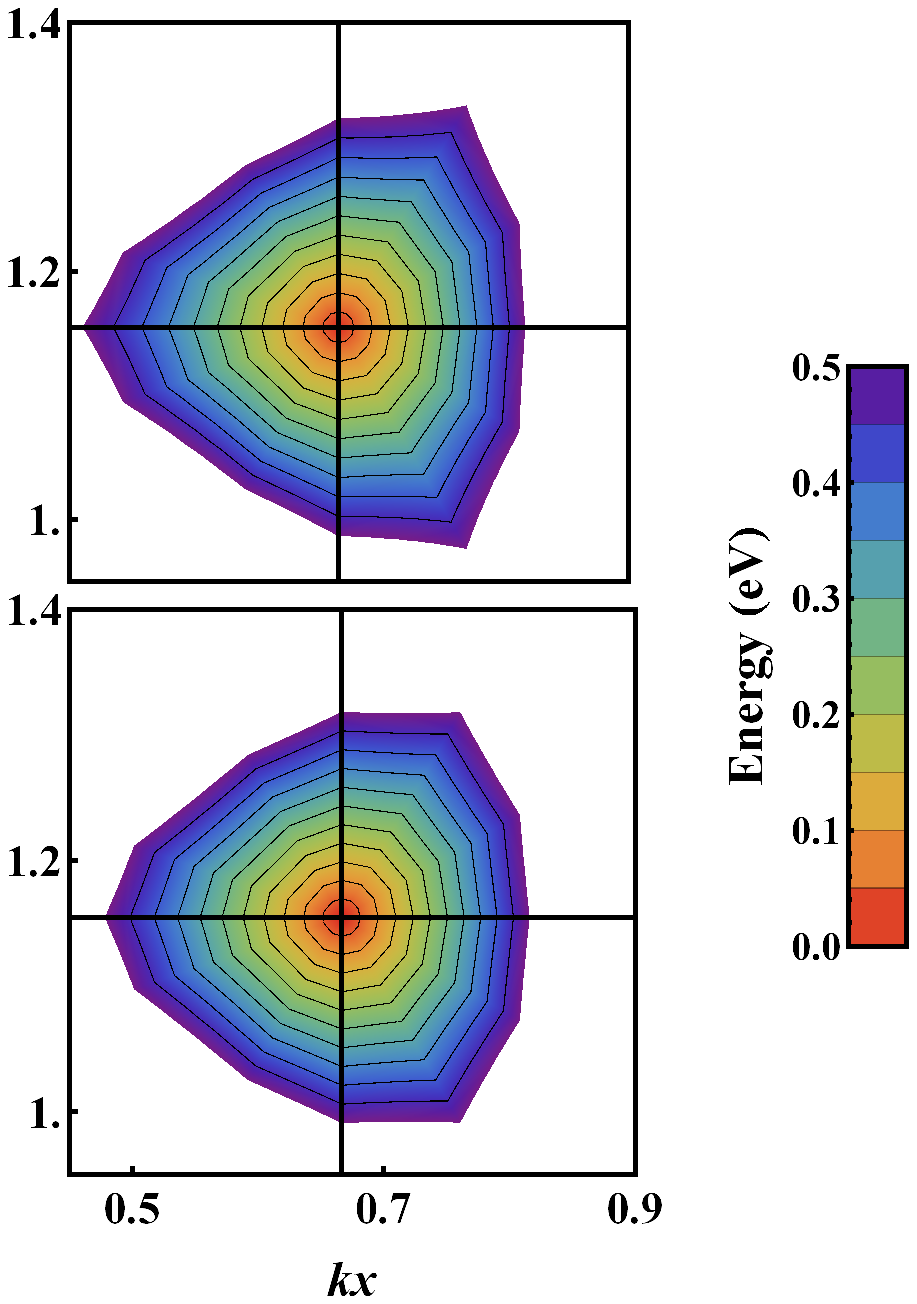}}\\
\caption{\small Hole (upper panels) and electron (lower panels) energy contour levels as a function of the wave vector: (a) h-Si and (b) h-Ge. Color online.}
\label{contours_h}
\end{figure}

Dirac cones do not display a circular shape.
This is more evident in a contour plot of energy levels in the neighborhood of the K-point (Fig. \ref{contours_h}).
The upper panels correspond to holes and lower panels to electrons.
Fig. \ref{contours_h} show that hole Dirac cones are  not only shorter than electron Dirac cones, but they are more asymmetric.
Again, this effect is more pronounced for h-Ge see right panels of Fig. \ref{contours_h}).

The contour plots reveal that the Fermi speeds along the K-M and K-$\Gamma$ directions are the smallest and largest Fermi speeds.
They also show that, in the neighborhood of the K-point, the energy diagram displays the symmetry of the dihedral group D$_3.$

\section{\label{sec:level5}CONCLUSIONS}

Our numerical studies of h-Si and h-Ge using DFT show their chemical stability.
We have reported their lattice parameters, densities of states and electronic band structures.
Their band structures, in particular, display Dirac cones in the vicinity of their K-points.
These cones display two types of asymmetries (deviation for circular shape and electron-hole asymmetry), which hint towards physics beyond the usual tight binding model of honeycomb lattices.
Fermi speeds are direction-dependent (even for graphene) because Dirac cones are not circular (in fact, they exhibit the symmetry of the equilateral triangle).
Hole Dirac cones are shorter and more asymmetric than electron Dirac cones.
Spin effects contribute to these asymmetries.
Due to  the two type of asymmetries described above, h-Si and h-Ge can not be accurately modeled by a Dirac equation for massless fermions, which predict direction-independent Fermi speeds ($E_{\pm}(\vec{k}) \approx \pm {v}_F\|\vec{k}\|,$ \cite{RevModPhys.81.109}), and a variant of Dirac equation, which incorporates the dependency on $k$-direction, must be sought.

\begin{acknowledgments}
Authors would like to thank the financial support by Division de Investigaci\'on Sede Bogot\'a, Universidad Nacional de
Colombia, (DIB), under Project 12584.
\end{acknowledgments}

\end{document}